\documentclass{PoS}
\usepackage{amstext}
\usepackage{amsmath}
\usepackage{fontenc}
\usepackage{times}
\usepackage{mathptmx}
\usepackage{graphicx}

\newcommand{\be}{\begin{equation}}
\newcommand{\ee}{\end{equation}}
\newcommand{\ba}{\begin{eqnarray}}
\newcommand{\ea}{\end{eqnarray}}

\title{Quantization of the  Myers-Pospelov model: a progress report}

\ShortTitle{Quantization of the  Myers-Pospelov model}

\author{C. M. ~Reyes, \speaker{~L. ~Urrutia} ~and ~J. D. ~Vergara%
         \\
        Universidad Nacional Aut\'onoma de M\'exico-MEXICO\\
        E-mail: \email{carlos.reyes@nucleares.unam.mx, urrutia@nucleares.unam.mx, vergara@nucleares.unam.mx} }

\abstract{The Myers-Pospelov (MP)  model is an effective field
theory (EFT), including dimension five operators, which describes
the phenomenology of active Lorentz invariance violation produced
by a preferred reference frame. We concentrate here in the case of
the modified electrodynamics. The point of view taken in this work
is that the Lorentz violating part of the action in the MP model,
which includes higher order time derivative (HOTD) operators, is
to be considered as a perturbation over the dynamics described by
standard Electrodynamics, particularly in the quantum case. HOTD
theories, besides incorporating additional degrees of freedom,
suffer from well known difficulties in their quantization, among
which one finds Hamiltonians which are not bounded from below.
Thus, in order to cope with these challenges it will be necessary
to deal with a modified perturbation theory which is well
described in the literature. We apply such methods to this
specific model providing a quantization of the free sector of the
theory by identifying the modified normal mode expansions of the
fields together with  the modified propagators and the interaction
terms. The calculation of interacting processes, together with
radiative corrections, is beyond the scope of the present article
and will be deferred for future publications.}

\FullConference{From Quantum to Emergent Gravity: Theory and Phenomenology\\
        June 11-15 2007\\
        Trieste, Italy}

\begin{document}

\section{Introduction}
The MP model \cite{MP}\ is an EFT theory that incorporates
scalars, fermions and photons in an observer (active) Lorentz
violating theory using dimension five operators together with the
presence of a fixed direction selecting a preferred frame. It has
recently been generalized to a non-abelian model including
interactions arising from the fields associated to the Standard
Model \cite{BP}, thus providing a dimension-five-operator
generalization of the original Standard Model Extension
\cite{KOSTELECKY1}.

In this work we will concentrate upon the simpler version of
Ref.\cite{MP}, particularly upon the proposed modified
electrodynamics. We present here the first steps to our final
goal, which is to provide a quantum version of MP electrodynamics
considered as a perturbation of standard QED, in the precise sense
that when making zero the parameters encoding the corrections we
must recover the well known results for any physical  process in
Lorentz covariant QED. This requirement is motivated by the fact
that all experimental and observational evidence point to
negligible Lorentz invariance violation at standard model
energies. The amazingly precise experimental predictions of
standard QED can be obtained as the result of perturbative
calculations, which now will require the incorporation of the
extra perturbation arising from  LIV into the scheme. An
additional challenge arises because the presence of dimension five
operators encoding the corrections to QED in the MP model make the
theory of the HOTD type. At least at the perturbative level, it is
well known that HOTD theories, besides the obvious property of
having additional degrees of freedom with respect to the lower
order ones, give rise to Hamiltonians which are not positive
definite, irrespectively of the interaction terms
\cite{JAEN,SIMON}. In fact, a perturbation of electrodynamics
should not introduce additional degrees of freedom, so that a
careful strategy is required to define an adequate perturbative
procedure in the LIV parameters. Fortunately, a systematic
approach to carry out this task, which also  provides a positive
definite zeroth order Hamiltonian, already exists in the
literature \cite{ELIEZER,CHENGETAL} and here we adhere to it.

In view of the above considerations we will proceed in the
following way to define the quantum field theory extension of the
MP model: (i) as usual, our starting point will be the classical
version of it given in Ref.\cite{MP} and which has been thoroughly
studied in relation to synchrotron radiation in Refs. \cite{MU}.
(ii) next we apply the procedure of Ref.\cite{CHENGETAL}  to the
classical HOTD MP model and reduce it to a modified effective
theory of the same time derivative character as classical
electrodynamics. The procedure leads to field redefinitions plus
additional contributions to the interactions. (iii) finally we
take this resulting classical theory as the correct starting point
for quantization, which we carry along the standard lines. The
resulting quantum theory provides the basis for the  calculation
of interacting processes using the standard perturbative scheme of
 quantum field theory (QFT). In this sense it is clear that we
are not producing a quantum version of the full MP model, but only
one which is adapted to our basic requirement of describing the
LIV corrections as perturbations to QED.

Perhaps we should emphasize at this stage that we are dealing with
two different classes of perturbations: the first one concerns
only the LIV parameters, occurs at the classical level and serves
to define the correct starting point for quantization. Once the
resulting  theory is quantized, the usual  QFT interacting
processes can be calculated, corresponding to the second class of
perturbations. Both approximations should be made consistent when
predicting a result to a given order in any of the LIV parameters.

Furthermore, in this work we only deal with the calculation of the
free propagators, which nevertheless incorporate modified
dispersion relations to lowest order in the LIV  parameters. In
other words, we identify the corresponding free propagating
excitations which will subsequently subjected to interactions. The
calculation of interacting  processes either at the tree or at the
one loop level are deferred to future publications. Of particular
interest to us will be the calculation of self-energies which can
be used as first indications of the fine-tuning problems arising
in some LIV theories \cite{CPSUV, GV}.

\section{The MP electrodynamics}
The corresponding free field Lagrangian densities are given by%
\begin{eqnarray}
\mathcal{L}_{photon} &=&-\frac{1}{4}F_{\mu \nu }F^{\mu \nu }+\frac{\xi }{M}%
E_{i}\partial _{0}B_{i},  \label{PHOTON} \\
\mathcal{L}_{fermion} &=&\;\;\bar{\Psi}i\gamma ^{\mu }\left( \partial _{\mu
}-m\right) \Psi +\frac{1}{M}\bar{\Psi}\gamma ^{0}\left( \eta _{1}+\eta
_{2}\gamma _{5}\right) \partial _{0}^{2}\Psi,  \label{FERMION}
\end{eqnarray}
in the particular frame where the Lorentz symmetry is broken in
the direction $n^{\mu }=(1,\mathbf{0})$ and in standard notation with metric $%
(1,-1,-1,-1)$. We interpret the scale $M$, together with the
dimensionless parameters $\xi ,\eta _{1} $  and $\eta _{2}$, as
the effective low energy imprints upon standard particle dynamics
produced by a fundamental quantum gravity theory, which has
induced an spontaneous LIV characterized by the vacuum expectation
value $n^{\mu }$. The choice $M=M_{Planck}$ leads to the
observational/experimental  bounds $|\xi| < 10^{-7}$ and
$|\eta_{\pm}=\eta_1 \pm \eta_2| < 10^{-5}$ \cite{MATTINGLY}. In
the limit of such parameters going to zero we demand that the
standard Lorentz covariant quantum results for electrodynamics are
recovered. After some basic features of the theory are revealed we
present a more detailed discussion of the relevant energy scales
that define the effective model in the last section.

In order to assess the real character of a HOTD contribution it is
convenient to look at the contribution of the Lorentz violating
terms to the equations of motion. The fermion modification will
add a second-order time derivative to the standard first-order
equation of motion. Additional degrees of freedom will appear in
this case. On the contrary, the photon contribution only
incorporates a modification to the standard second-order term in
electrodynamics and no additional degrees of freedom are present.
Nevertheless, such modifications will still require some field
redefinitions in order to exhibit a positive definite Hamiltonian
consistent with the field operators commutation relations.
\section{The perturbative expansion}
The general method for dealing with the canonical description of
HOTD theories was given a long time ago in \ Ref.
\cite{OSTROGRADSKI}. In order to highlight some general features
of these theories we briefly review their basic properties in the
context of a scalar field theory satisfying the non-degeneracy
condition $\left(\frac{{\partial}^2\mathcal{L}}{\partial
\phi^{(k)}\partial \phi^{(k)}}\right)\neq 0$,\ where the fields
depend upon the space-time coordinates. The generalizations
incorporating spinor and
vector fields they have been analyzed in Refs. \cite{ELIEZER},\ \cite%
{CHENGETAL1}.

If the highest time derivative in the Lagrangian density  $%
\mathcal{L}=\mathcal{L}(\phi(t,\mathbf{x}),...,\phi^{(k)}(t,\mathbf{x}))$,
is of order $k$, with the notation  \footnote{For the lower order
time derivatives we might use also the standard dots notation}
\begin{equation}
\phi^{(k)}(t,\mathbf{x})=\frac{\partial^{k}\phi(t,\mathbf{x})}{\partial t^{k}%
},
\end{equation}
 the corresponding phase space will be of
dimension $2k$ per space
point, been characterized by$\;k\;$ fields $:Q_{0}=\phi(t,\mathbf{x}%
),\;Q_{1}=\phi^{(1)}(t,\mathbf{x}),....,\;Q_{k-1}=\phi^{(k-1)}(t,\mathbf{x}%
)\;$together with $k\;$momenta
\begin{equation}
P_{i}(t,\mathbf{x})=\frac{\partial \mathcal{L}}{\partial \phi^{(i+1)}} +
\sum_{j=1}^{k-i-1}\left(-\frac{\partial}{\partial t} \right)^j \frac{%
\partial \mathcal{L}}{\partial \phi^{(j+1)}}, \quad i=0, \dots, (k-1).
\label{GENMOM}
\end{equation}%
Here and in the sequel, as far as no confusion arises, we avoid writing the
explicit space-time dependence in the fields. The equation of motion for $%
\phi\;$will be of order $2k$ in the time derivatives, requiring the fixing
of $2k\;$initial conditions, which is consistent with the existence of $2k\;$%
degrees of freedom phase space. The Hamiltonian is%
\begin{equation}
H=\int d^3x \left( \sum_{i=0}^{k-1}\;P_{i}Q_{i}-\;\mathcal{L}%
(Q_{0},\dots,Q_{k-1},\;\phi^{(k)}(P_{k-1},Q_{0},...,Q_{k-1}))
\right), \label{HAMHOTD}
\end{equation}
where we have used the non-degeneracy condition which implies that we can
solve $\phi^{(k)}\;$as a function of $Q_{0},...,Q_{k-1},P_{k-1}.\;$The above
expression is linear in the momenta $P_{i},\;i=0,...,k-2$, thus making the
Hamiltonian (\ref{HAMHOTD}) unbounded from below, independently of the
interaction terms included in the Lagrangian.

Since we are interested in dealing with HOTD corrections in the
action as perturbations upon standard theories we must rely on a
procedure which (i) retains the original number of degrees of
freedom and (ii) produce free Hamiltonians bounded from below as
adequate starting points for quantization. Such a method has been already developed in Refs. %
\cite{ELIEZER,CHENGETAL} and we present here a brief summary of it
adapted to the case of a scalar field theory. In order to point
out some its basic features let us consider the simplest framework
of a non-covariant Lagrangian density, analogous to those
described in Eqs. (\ref{PHOTON}), (\ref{FERMION}), depending upon
accelerations and where the HOTD contribution is only present as a
perturbation characterized by the small parameter $g$
\begin{equation}
\mathcal{L}(\phi,\partial_\mu{\phi},\ddot{\phi})=\mathcal{L}%
(\phi,\partial_\mu{\phi})+ \frac{1}{2}\, g \, \ddot{\phi}^2.
\end{equation}
In this case the quantities $\phi, \dot{\phi} $ play the role of coordinate
fields. The standard procedure of extremizing the action leads to
\begin{eqnarray}
\delta S&=&\delta \int d^4x \mathcal{L}(\phi, \partial_\mu{\phi}, \ddot{\phi}%
)= \int d^4x \ \partial_\mu \left[\left(\frac{\partial \mathcal{L}}{%
\partial\partial_\mu \phi}-\partial_\nu\left( \frac{\partial \mathcal{L}}{%
\partial\partial_\mu\partial_\nu \phi}\right)\right) \delta \phi \right.
\notag \\
&&+\left. \frac{\partial \mathcal{L}}{\partial \partial_\mu\partial_\nu\phi}%
\delta\partial_\nu{\phi}\right] +\int d^4x \,\,
E\left([\phi]\right)\delta \phi, \label{VARACT}
\end{eqnarray}
where
\begin{equation}
E\left([\phi]\right)=\partial_\mu\partial_\nu\left(\frac{\partial \mathcal{L}%
}{\partial\partial_\mu\partial_\nu \phi}\right)-\partial_\mu \left(\frac{%
\partial \mathcal{L}}{\partial \partial_\mu \phi}\right)+ \frac{\partial
\mathcal{L}}{\partial \phi}=0.  \label{EQSMOT}
\end{equation}
The first term in the right hand side of Eq. (\ref{EQSMOT}) gives
the fourth order time derivative contribution $\phi^{(4)}(t,
{\mathbf x})$ to the equation of motion. The space-time dependent
momenta, associated to $\phi$ and ${\dot \phi}$ respectively, can
be directly read off from the surface boundary term in Eq.
(\ref{VARACT})
\begin{equation}
P_0= \frac{\partial \mathcal{L}}{\partial \dot{\phi}}-\frac{\partial}{%
\partial t}\left(\frac{\partial \mathcal{L}}{\partial \ddot{\phi}}\right),
\quad P_1= \frac{\partial \mathcal{L}}{\partial \ddot{\phi}},
\end{equation}
in accordance with the general expression in Eq. (\ref{GENMOM}).  From the
simple form assumed for the HTOD term, which satisfies the non-degeneracy
condition,  it is clear that both velocities can be solved in terms of the
momenta: $\ddot{\phi}$ can be expressed in terms of $P_1$, and $\phi^{(3)}$
in terms of $P_0$. Nevertheless, notice that  both substitutions carry the
non-analytical factor $1/g$. This  is precisely what makes non-trivial a
perturbative expansion around $g=0$.

The full Hamiltonian $H$ and symplectic form $\Omega$ are defined
according to the Ostrogradski  procedure as
\begin{equation}
H=\int d^3x \left(P_0 \dot{\phi}+P_1 \ddot{\phi}- \mathcal{L}\right), \qquad
\Omega =\int d^3x \ d^3y \left(dP_0(t,\mathbf{x}) \wedge d {\phi(t, \mathbf{%
y })}+ dP_1(t,\mathbf{x}) \wedge d {\dot \phi(t,\mathbf{y})}\right).
\end{equation}
The dangerous contributions to the Hamiltonian arise from the non-analytic
term $P_1^2/2g$ together with the unbounded piece $P_0\, \dot{\phi}$.

Let us summarize now the general perturbative procedure for the
non-degenerate case according to the method in
Ref.\cite{CHENGETAL}, in the framework of a system having a
Lagrangian density of the form
\begin{equation}
\mathcal{L}=\mathcal{L}_0(\phi \, , \, \partial_\mu{\phi})+g \mathcal{L}%
_1(\phi \, , \, \partial_\mu{\phi}, \, \phi^{(2)} , \, \dots \, , \,
\phi^{(n)}),
\end{equation}
where $g$ is a small parameter. The steps are the following: (i)
in order to obtain the appropriate Hamiltonian to order $g^k$, one
starts by iteratively solving the equations of motion to order
$g^{(k-1)}$.
(ii) next express all time derivatives $\phi^{(k)}(t,%
\mathbf{x})$ $\ $for $k > 2$ in terms of the lowest
time-derivative fields describing the unperturbed system,
which  are $\phi (t,\mathbf{x})\;$and $%
\dot{\phi}(t,\mathbf{x})$ in our example.  This will introduce
further contributions in powers of the perturbation parameters
which need to be maintained only up to the required order. (iii)
then rewrite the Hamiltonian together with the symplectic form
obtained from the Ostrogradski method by substituting  the momenta
together with all $\phi ^{(k)}(t,\mathbf{x}),\,\, k > 2$ in terms
of these basic variables.
\begin{equation}
\mathcal{H}=\mathcal{H}(\phi, \, \dot{\phi}, \nabla\phi),\qquad \Omega=\int
d^3 x \ d^3y \ \omega(\phi(t, \mathbf{x} ), \, \dot{\phi}(t, \mathbf{y}))\, d%
\dot{\phi}(t, \mathbf{y})\, \wedge\, d \phi(t, \mathbf{x}) ,
\end{equation}
from where we can read the bracket $\{ \phi(t, \mathbf{x}) \, , \, \dot{\phi}%
(t, \mathbf{y})\}$. (iv) finally find an invertible change of
variables $\phi (t,\mathbf{x}),$
$\dot{\phi}(t,\mathbf{x})\rightarrow Q(\phi
,\dot{\phi},...),\;P(\phi ,\dot{\phi},...)$ in such a way that the
corresponding Poisson bracket $\{Q(t,\mathbf{x})\, , \, P(t,%
\mathbf{y})\}$ is canonical to the order considered. That is to
say
\begin{eqnarray}
&&\delta^3 ({\mathbf x}- {\mathbf y}) + O(g^{k+1})=\{Q(t,\mathbf{x})\, , \, P(t,%
\mathbf{y})\}\notag \\
&&= \int d^3z \ \ d^3z^{\prime}\left(\frac{\delta
Q(t,\mathbf{x})}{\delta
\phi(t,\mathbf{z})}\frac{\delta P(t,\mathbf{y})}{\delta \dot{\phi}(t,\mathbf{%
z^{\prime}})}- \frac{\delta Q(t,\mathbf{x})}{ \delta \dot{\phi}(t,\mathbf{%
z^{\prime})}}\frac{\delta P(t,\mathbf{y})}{\delta \phi(t,\mathbf{z})}%
\right)\{ \phi(t,\mathbf{z}) \, , \,
\dot{\phi}(t,\mathbf{z^{\prime}})\}.
\end{eqnarray}
At last, the Hamiltonian density $\tilde{\mathcal{H}}(Q,P, \dots )=\mathcal{H%
}(\phi(Q, P, \dots)\, , \, \dot{\phi}(Q,P, \dots))$ together with the
Poisson bracket $\{Q (t,\mathbf{x}), P(t,\mathbf{y})\}=\delta^3(\mathbf{x}-%
\mathbf{y})$ define the physical approximation of the system to the order
considered. This Hamiltonian will be bounded from below provided the initial
one obtained from $\mathcal{L}_0$ is. The effective Lagrangian density is
given by $\tilde{\mathcal{L}}(Q\, , \dot{Q},\dots)= P\, \dot{Q}-\tilde{%
\mathcal{H}}$ and the quantization is straightforward since it is
first order.

A proof of self-consistency to all orders, in a mechanical
setting\footnote{Field theory in $1+{\mathbf 0}$ dimensions}, is
provided in Ref.\cite{CHENGETAL}. It consists in showing that to
each order in the expansion parameter, the Lagrangian constructed
according to the summary described in the above paragraph
reproduces exactly the corresponding equations obtained by the
iteration procedure stating from the exact HOTD ones.

An illuminating example of the relevance of the above procedure to
our case is also given in Ref. \cite{CHENGETAL}. The authors
consider a system of two one-dimensional  oscillators coupled in the
presence of a constant gravitational field. Each oscillator has
natural free frequencies $\Omega _{0}=\sqrt{{K}/{M}}$ and $\omega
_{0}=\sqrt{{k}/{m}}$, respectively. The full normal modes
frequencies $\Omega $ and $\omega$ can be exactly calculated. On the
other hand, the equations of motion consisting in two coupled second
order differential equations can be uncoupled through a fourth order
differential equation for one coordinate, which can be obtained from
an  acceleration dependent Lagrangian. Next, they look at the
situation in terms of a perturbative scheme starting from this
acceleration dependent Lagrangian.  The expansion parameter is taken
to be $g={k}/{K} \ll 1 $ and for the sake of the discussion the two
masses are considered of the same order, i.e. $m\sim M$. When
probing energies of the order of $\omega _{0}$ one verifies that the
acceleration corrections in the Lagrangian become negligible with
respect to the velocity and coordinate dependents ones, which
coefficients depend upon all the parameters. The most immediate
possibility to proceed along the lines of maintaining the original
number of degrees of freedom $x,\dot{x}$, is to simply neglect the
acceleration term and compute the corrections to $\omega$ to first
order in $g$. Nevertheless, this result does not coincide with the
first order expansion of the exact frequency $\omega$ to that order.
On the other hand, the application of the modified perturbative
method proposed in \cite{CHENGETAL} does indeed leads to the correct
expression to that order. Moreover, in this simple case, the
corrections can be calculated to all orders in $g$ and the sum of
this perturbation series can also be performed, leading to the exact
expression for $\omega$. The high frequency mode $\Omega $, which is
non-analytical  when written in terms of the parameter $g$, is not
seen by the procedure, meaning that the results are valid only for
energies much lower than $\Omega$. One could summarize the above
description by saying that the physical meaning of reducing the
enlarged original configuration $
\left({x,\dot{x}}\right)$ - velocity (${\ddot{x},%
\dddot{x}}$) space  to that generated by $x$ and $\dot{x}$ in the
proposed perturbative formulation is to allow the calculation of
further corrections to the excitations of the low energy modes
already present in the zeroth order system, in a way consistent
with the exact evolution. We refer the reader to the original
paper for further details.

\section{The perturbative expansion in the Myers-Pospelov model}
Here we describe the main ingredients and results of the application
of the method described in the previous section to the quantization
of the free sector in MP electrodynamics. Our general strategy will
be the following: (i) since the fermions acquire corrections of the
HOTD type, we start by constructing the corresponding effective
Lagrangian and Hamiltonian densities to a desired order in $g=1/M$.
Then we quantize this effective theory and calculate the modified
free propagator. (ii) next we introduce the photons via minimal
coupling in this effective fermion Lagrangian density and also add
the contributions of Eq. (\ref{PHOTON}), identifying the resulting
free and interaction terms. (iii) subsequently we find the correct
effective Hamiltonian formulation for the free photon field and
proceed to its quantization, obtaining also the modified propagator.
In both cases the normal modes of the free sector correspond to
particles with propagation properties which are different from the
usual ones described by the limit $ \eta_1/M, \eta_2/M \rightarrow
0$.

\subsection{The fermionic sector}

Here we consider corrections to order $g=1/M >0 $ and start from
the HOTD Lagrangian density \be
{\cal L}=\bar{\psi}(i\gamma^{\mu }\overrightarrow{\partial_{\mu }}%
-m)\psi +g \, \bar{\psi}\Sigma_0\ddot{\psi}, \quad \Sigma_0=
\gamma^{0}\left( \eta _{1}+\eta _{2}\gamma _{5}\right), \ee which
produces the equation of motion \be
\dot{\psi}=\overrightarrow{\alpha}\psi +ig\chi \ddot{\psi}, \quad
\overrightarrow{\alpha}=-\gamma _{0}(\gamma
^{i}\overrightarrow{\partial _{i}}+im). \label{OPALFA} \ee In the
approximation $\dot{\psi}=\overrightarrow{\alpha}\psi,
\,\ddot{\psi}=\overrightarrow{\alpha}\dot{\psi}$, the canonical
momenta, the Hamiltonian density, and the symplectic form,
respectively, are \ba \Pi
_{0\psi }&=&\bar{\psi}i\gamma ^{0}+g\bar{\psi}\overleftarrow{%
\alpha }\Sigma _{0},\quad \Pi _{1\psi }=g\bar{\psi}\Sigma _{0},
\quad  \Pi _{0\bar{\psi}} = 0 = \Pi _{1\bar{\psi}},\\
\mathcal{H}&=&-i\bar{\psi}\gamma ^{k}\partial _{k}\psi +m\bar{\psi}\psi +g%
\bar{\psi}\overleftarrow{\alpha }\Sigma _{0}\overrightarrow{\alpha
}\psi, \\
\Omega &=&\int d^{3}x\left\{ id\psi ^{\dag }\left[ 1-ig\gamma ^{0}%
\overleftarrow{\alpha }\Sigma _{0}-ig\gamma ^{0}\Sigma _{0}\overrightarrow{%
\alpha }\right] \wedge d\psi \right\}. \label{SIMPLECTIC}\ea The
required change of variables $\psi \rightarrow \tilde{\psi}$ to
recover the standard symplectic structure \be
\{\tilde{\psi}_{A}(t,\mathbf{x}),\tilde{\psi}_{B}^{\dag }(t,\mathbf{y}%
)\}=\delta _{AB}\delta ^{3}(\mathbf{x}-\mathbf{y}), \ee where the
labels $A,B=1,2,3,4$ denote the four-spinor components, is  \be
\tilde{\psi}=(1-ig\gamma _{0}\Sigma _{0}\overrightarrow{\alpha
})\,\psi, \qquad \bar {\tilde{\psi}}={\bar{\psi}}\,(1-ig%
\overleftarrow{\alpha }\Sigma _{0}\gamma ^{0}). \ee This leads to
the following effective Lagrangian and Hamiltonian densities  \ba
\tilde{\mathcal{L}}&=&\bar {\tilde{\psi}}\left( i\gamma ^{\mu
}\partial _{\mu
}-m\right) \tilde{\psi}+g\bar {\tilde{\psi}}\gamma _{0}\left( \overrightarrow{%
\alpha }\right) \gamma^0 \Sigma_0 \left( \overrightarrow{\alpha
}\right) \tilde{\psi}, \label{EFFLAG}\\
\tilde{\mathcal{H}} &=& {\tilde{\psi}}^{\dag
}\left[ -i \overrightarrow{\alpha} \tilde{\psi}%
-g(\gamma _{0}\overrightarrow{\alpha }\chi \overrightarrow{\alpha })\tilde{%
\psi}\right]= {\tilde \psi}^\dag i \frac{\partial {\tilde \psi}
}{\partial t}. \label{EFFHAM}\ea Next we proceed to quantization
by introducing fermionic creation (annihilation) operators for
particles $b^\dagger_\lambda (\mathbf{k}), (b_\lambda
(\mathbf{k}))$ and for antiparticles $d^\dagger_\lambda
(\mathbf{k}), (d_\lambda (\mathbf{k}))$, with standard
anticommutation relations. The label $\lambda= \pm 1$ denotes the
helicity quantum number. The normal modes (physically propagating
particles) corresponding to the full Dirac equation derived from
Eq.(\ref{EFFLAG}) are characterized by  positive energies $
E_u^{\lambda}({\mathbf k}), E_v^{\lambda}({\mathbf k}) $ according
to the modified dispersion relations  \begin{eqnarray}
E_{u}^{\lambda }({\mathbf k}) &=&E_{0}+g(\eta
_{1}E_{0}^{2}+\lambda \left| {\mathbf k}\right| \eta _{2}E_{0}),
\quad  E_{v}^{\lambda }({\mathbf k}) =E_{0}-g(\eta
_{1}E_{0}^{2}+\lambda \left| {\mathbf k}\right| \eta _{2}E_{0}),
\label{PHYSNEGEN1}
\end{eqnarray}
where the notation is $E_{0}=+ \sqrt{{\mathbf k}^2+m^2}$. Here we
can appreciate one indication that we are dealing with an EFT: it is
clear that for any choice of the parameters $\eta_1, \eta_2$, in the
high momentum limit there will be at least one value of the above
energies which will be negative, thus producing a Hamiltonian which
is not positive-definite. To avoid this possibility we must impose
the restriction $|\mathbf{k}|< M/|\eta_1+\lambda\, \eta_2|$. Further
discussion of this issue is given in the last section.

In terms of the above normal modes, the fermionic field can be
expanded as
\begin{eqnarray}
\tilde{\psi}_{A}(x)=\int \frac{d^{3}{{\mathbf k}}}{\sqrt{(2\pi )^{3}}} %
\,\sum_{ \lambda=\pm  } \sqrt{\frac{m}{E_{u}^{\lambda }({\mathbf{k}})}}%
\,b_{\lambda }({\mathbf{k}})\,U_{A\lambda
}({\mathbf{k}})e^{-ik_{u}^{\lambda }\cdot x} \nonumber \\ +
\int \frac{d^{3}{\mathbf{k}}}{\sqrt{(2\pi )^{3}}} %
\,\sum_{\lambda=\pm  }
\sqrt{\frac{m}{E_{v}^{\lambda }({\mathbf{k}})}}d_{\lambda }^{\dag }(%
{\mathbf{k}})\,V_{A\lambda }({\mathbf{k}})e^{ik_{v}^{\lambda
}\cdot x}.\label{FIELDEXP}
\end{eqnarray}
Here the notation is $\, k_{u}^{\lambda }\cdot
x=E_u^\lambda\,t-{\mathbf k}\cdot {\mathbf x}$ and analogously $\,
k_{v}^{\lambda }\cdot x=E_v^\lambda\,t-{\mathbf k}\cdot {\mathbf
x}$. Also, $U_{A\lambda }({\mathbf{k}})$ and $V_{A\lambda
}({\mathbf{k}})$ are the corresponding eigenspinors of the one
particle Hamiltonian associated to (\ref{EFFHAM}). Such spinors
can be explicitly written to order $g$ and satisfy the following
properties \ba V_{\lambda }(g,{\mathbf k})&=&\gamma _{5}U_{\lambda
}(-g,{\mathbf k}),\qquad V_{\lambda }^{\dag }({\mathbf
k})U_{\lambda
^{\prime }}({-{\mathbf k}})=0,\label{PROP1}\\
U_{\lambda }^{\dag }({\mathbf k})U_{\lambda ^{\prime }}({\mathbf
k})&=&\delta ^{\lambda \lambda ^{\prime }}\frac{E_{u}^{\lambda
}({\mathbf k})}{m}, \qquad V_{\lambda }^{\dag }({\mathbf
k})V_{\lambda ^{\prime }}({\mathbf k}) =\delta ^{\lambda
\lambda ^{\prime }}\frac{E_{v}^{\lambda }({\mathbf k})}{m}, \label{PROP2}\\
\frac{m}{E_{u}^{\lambda }}U_{\lambda }({\mathbf
k})\bar{U}_{\lambda }( {\mathbf k})&=&\frac{1}{2E_{0}}\left(
\gamma_0 E_0-{\mathbf \gamma}\cdot {{\mathbf k}} +m-igm\eta
_{2}\gamma
_{5}\alpha _{-{\mathbf k}}\right) P^{\lambda} \label{SUMU}\\
\frac{m}{E_{v}^{\lambda }}V_{\lambda }({\mathbf
k})\bar{V}_{\lambda }( {\mathbf k})&=&\frac{1}{2E_{0}}\left(
\gamma_0 E_0-{\mathbf  \gamma}\cdot {{\mathbf k}}-m+igm\eta
_{2}\gamma _{5}\alpha _{{\mathbf k}}\right) P^{\lambda
}.\label{SUMV} \ea Here $\alpha _{{\mathbf k}}$ is the momentum
representation of the operator $\overrightarrow{\alpha}$ defined
in Eq. (\ref{OPALFA}) and $P^\lambda$ is the helicity projector
\be
P^{\lambda} =\frac{1}{2}\left( I  + \lambda \frac{{{ \Sigma} \cdot {\mathbf k}}}{|{\mathbf k}|}%
\right), \ee where ${\Sigma}$ is the spin operator.

Using  the field expansion (\ref{FIELDEXP}) together with the
required  properties from Eqs. (\ref{PROP1}), (\ref{PROP2}), the
Hamiltonian, the momentum and the charge operators have the
expected form in terms
of the number operators $ b_{\lambda }^{\dag }({\mathbf k}%
)b_{\lambda }({\mathbf k})$ and $d_{\lambda }^{\dag }({\mathbf k})d_{\lambda }(%
{\mathbf k}).$ Our final task in this subsection is to write the
fermion propagator \be iS_{AB}(x-y)=\langle 0|T(\psi
_{A}(x)\,\bar{\psi}_{B}(y))|0\rangle, \ee which, in momentum
space, is given by \begin{eqnarray}
 S(k_0, {\vec k}) = \, \sum _{\lambda= \pm } \left[ \frac{1}{(k_0- {E}_u^{\lambda}
+i\epsilon)} \left\{\frac{m} {E_u^{\lambda}}U_{\lambda}(\vec{k})
\bar U_{\lambda} (\vec{k})\right\}
 + \frac{1}{(k_0+ {E}_v^{\lambda}-i\epsilon) } \left\{\frac{m} {%
E_v^{\lambda}}V_{\lambda}(-\vec{k}) \bar V_{\lambda}(%
-\vec{k})\right\} \right].\nonumber \\
\end{eqnarray}
Let us emphasize that the terms in curly brackets appearing in the
above expression have been explicitly calculated in Eqs.
(\ref{SUMU}) and (\ref{SUMV}). Also the poles in $k_0$ appear as
exact functions of the normal mode energies to the order
considered.

\subsection{The photon sector}

In order to include  photon field $A_\mu$ we perform the minimal
substitution $\partial_\mu \rightarrow (\partial_\mu+ieA_\mu)$ in
the effective Lagrangian density (\ref{EFFLAG}), to which  we add
the free contributions from  (\ref{PHOTON}). The result is
\begin{eqnarray} \mathcal{L}_{QED} &\mathcal{=}&\bar
{\tilde{\psi}}\left( i\gamma ^{\mu }\partial _{\mu }-m\right)
\tilde{\psi}+g\bar{\tilde{\psi}}\gamma _{0}\left(
\overrightarrow{\alpha }\right)\gamma_0 \Sigma_0 \left(
\overrightarrow{\alpha }\right)
\tilde{\psi} +\frac{1}{2}\left( \dot{A}^{i}-\partial ^{i}A^{0}\right) ^{2}-\frac{1}{4}%
F_{ij}F^{ij}+{\bar g} \, \epsilon _{ijk}\dot{A}^{i}\partial _{j}\dot{A}^{k}\nonumber  \\
&&-\left(e\, \bar{\tilde{\psi}}\gamma
^{0}\tilde{\psi}\right)A_{0}+ {\cal L}_{int} (A^i, {\tilde
\psi}),\label{LAGFERMFOT}
\end{eqnarray}
with the standard definition for the electromagnetic tensor
$F_{\mu\nu}$. Here have introduced the new perturbation parameter
${\bar g}=\xi/M$. As we can see from (\ref{LAGFERMFOT}) the
additional photon contribution is not of the HOTD type.
Nevertheless, the correction term proportional to ${\bar g}$ in
Eq.(\ref{LAGFERMFOT}) will demand some modifications in the
correct quantization procedure. In the sequel we denote $e\,
\bar{\tilde{\psi}}\gamma ^{0}\tilde{\psi}=J^0$. The contribution
${\cal L}_{int} (A^i, {\tilde \psi})$, defining the interaction
among fermions and transverse photons,  has a linear and quadratic
dependence upon $A_i$ arising from the operator $\left( \overrightarrow{%
\alpha }\right) \gamma^0 \Sigma_0 \left( \overrightarrow{\alpha
}\right)$ in (\ref{EFFLAG}). Its precise form is not relevant now
for our purpose of quantizing the free sector of the theory.

In order to identify the proper normal modes of the photon it is
convenient to proceed according to the Hamiltonian formulation,
which goes along similar lines that in the standard case. Our
notation here corresponds to that of a  three-dimensional
euclidean space where the relevant  vectors are: ${\bf A}=(A^i),\,
{ {\Pi }}=(\Pi_i),\, {\bf \nabla}= (\partial_i), \, i=1,2,3,$  and
$\epsilon_{123}=1.$
 The corresponding canonically conjugated
momenta are \be
\Pi_0=0, \quad \Pi _{i}=\left( \delta_{ik}+2{\bar g} \epsilon _{ijk}\partial _{j}\right) \dot{A}%
^{k}+\partial_{i}A^{0}. \ee
 The photon sector of the
Hamiltonian density, calculated to order ${\bar g}$,  is
\begin{eqnarray}
\mathcal{H}_{QED, \gamma} &=&\frac{1}{2}\left( \Pi _{i}\right) ^{2}+\frac{1}{4}%
F_{ij}F^{ij}+\left( \partial _{i}\Pi _{i}+J^{0}\right)A^{0}-{\bar g} \, \epsilon _{ijk}\Pi _{i}\partial _{j}\Pi _{k}-\;\mathcal{L}%
_{int}(A^{i},{\tilde \psi}) \label{FOTMPED}.
\end{eqnarray}
The evolution of the constraint $\Pi_0=0$ produces the Gauss law
$\partial_i \Pi_i +J^0=0$ as a secondary constraint. Its further
evolution must be consistent with current conservation via the
equations of motion, so that  we recover  the standard two first
class constraints of electrodynamics. A first gauge fixing is
provided  by choosing \be \Pi_0=0, \quad
A^0=-\frac{1}{\nabla^2}(J^0+
\partial_0\partial_i A^i). \label{A0FIX}\ee
To complete the gauge fixing it is convenient to separate the
photon fields into longitudinal and transverse components \be \Pi
_{i}^{T} =\dot{A}_{T}^{i}+2{\bar g} \, \epsilon _{ijk}\partial
_{j}\dot{A}_{T}^{k}, \quad \Pi
_{i}^{L}=\dot{A}_{L}^{i}+\partial_{i}A^{0}, \quad
A_{L}^{i}=\frac{1}{\nabla ^{2}}\partial _{i}(\partial _{k}A^{k}),
\ee  and to demand the Coulomb gauge $\partial_i A^i=0$, which is
equivalent to the requirement of $A^i_L=0$. In this way the Gauss
law turns out to be identically satisfied in virtue of the
properties $\partial_i A^i=\partial_i A_L^i, \, \, \partial_i
\Pi_i=\partial_i \Pi^L_i$ together with choice (\ref{A0FIX}) for
$A^0$. The longitudinal fields are fixed according to {\be
A_{L}^{i}=0,\qquad \Pi _{i}^{L}=-\frac{1}{\nabla ^{2}}\partial_{i}
J^{0}. \ee In this way the remaining independent degrees of
freedom are the transverse fields \be
A^{i}=A_{T}^{i}, \qquad \Pi _{i}^{T}=\dot{A}_{T}^{i}+2{\bar g} \, \epsilon _{ijk}\partial _{j}\dot{A}%
_{T}^{k}, \label{TRANSV}\ee satisfying the Poisson brackets \be
\left\{ A_{T}^{i}(t,\mathbf{x}),\;\Pi _{m}^{T}(t, \mathbf{y})\right\} =\left( \delta_{im}-\frac{%
\partial_{i}\partial_{m}}{\nabla^{2}}\right) \delta
^{3}(\mathbf{x}-\mathbf{y}),\label{PBTF} \ee with the remaining
been zero.

In order to allow for consistency among: (i) the equal time
commutators  for the transverse fields arising from
Eq.(\ref{PBTF}), (ii) the expansion of them in term of frequency
modes and (iii) the standard bosonic creation-annihilation
operator commutation relations, we need to introduce a canonical
transformation (to order ${\bar g}$) which guaranties that
$\tilde{\Pi}_{i}^{T}=\partial _{0}\tilde{A}_{T}^{i}$, as opposed
to the second equation (\ref{TRANSV}). Such transformation is
\begin{eqnarray} \tilde{\Pi}_{i}^{T} =\left( \delta_{ik}-{\bar g}
\, \epsilon _{isk}\partial _{s}\right) \Pi _{k}^{T}, \qquad
\tilde{A}_{T}^{i} =\left( \delta_{ik}+{\bar g} \,  \epsilon
_{isk}\partial _{s}\right) A_{T}^{k},
\end{eqnarray}
which defines the physical fields of the theory.  The normal mode
expansion of the photon field  is \be
\tilde{A}_{T}^{i}(x)=\int \frac{d^{3}{\mathbf k}}{\sqrt{(2\pi )^{3}}}%
\,\sum_{\lambda= \pm }\sqrt{\frac{1}{2\omega _{\lambda }({\mathbf
k})}}\left[
\,a_{\lambda }({\mathbf k})\,\varepsilon ^{i}(\lambda ,{\mathbf k}%
)e^{-ik^{\lambda }\cdot x} \, + \,   h.c. \right]. \label{FOTEXP}
\ee Here $a_{\lambda }({\mathbf k}), a^{\dag}_{\lambda }({\mathbf
k}) $ are standard creation-annihilation operators and the complex
numbers $\varepsilon ^{i}(\lambda ,{\mathbf k})$ define  a
circularly polarized basis with helicity $\lambda$. The notation is
$k^{\lambda }\cdot x=\omega_{\lambda}({\mathbf k})\, t- {\mathbf
k}\cdot {\mathbf x}$, where the modified positive frequencies (to
first order in ${\bar g}$) are \be \omega_\lambda({\mathbf k})=
|{\mathbf k}|(1-\lambda {\bar g} |{\mathbf k}|).\label{FOTFREQ}\ee
Again, the effective character of the theory manifests itself in the
condition $|{\mathbf k}|< 1/{\bar g}=M/\xi$, which will be further
discussed in the last section.

In this way, the final photon Hamiltonian density arising from
(\ref{FOTMPED}) reads \begin{equation} {\tilde {\mathcal
H}}_{MP,\, \, \gamma}=\frac{1}{2}(\tilde{ \Pi}^{T})^2+ \left(
\frac{1}{2}{\tilde {\mathbf B}}^2  -{\bar g}\, {\tilde{\mathbf
B}}\cdot (\nabla \times {\tilde {\mathbf B}})\right)
-\frac{1}{2}J^0 \frac{1}{\nabla^2} J^0 -{\tilde{\cal
L}}_{int}({\tilde{A}_{T}}, {\tilde \psi} ).\label{FINHAMFOT}
\end{equation}
As usual ${\tilde {\mathbf B}}={\mathbf \nabla}\times {\tilde
{\mathbf A}}$. Let us emphasize that the contribution $\frac{1}{2}
{\tilde {\mathbf B}}^2 -{\bar g}\, {\tilde{\mathbf B}}\cdot
(\nabla \times {\tilde {\mathbf B}})=\frac{1}{2}
\left(\tilde{\mathbf B}-{\bar g}\, \nabla \times {\tilde {\mathbf
B}}\right)^2 $ is positive definite to order ${\bar g}$. This can
also be verified by calculating the normal ordered expression for
the free sector of the Hamiltonian arising from (\ref{FINHAMFOT}),
using the expression (\ref{FOTEXP}), which leads to the expected
result \be {\tilde H}_{\, 0}=\int d^3 {\mathbf k} \sum_{\lambda=
\pm}\, \omega_\lambda({\mathbf k})\, a_\lambda^{\dag}({\mathbf
k})\, a_\lambda({\mathbf k}), \ee in terms of the corresponding
positive frequencies (\ref{FOTFREQ}), to the order considered.

The transverse photon propagator is
\begin{equation} i\Delta_T^{ij} (x,y)
\equiv \left\langle 0 \right|T\left( {{\tilde A}^i_T \left( x
\right){\tilde A}^j_T \left( y \right)} \right)\left| 0
\right\rangle.
\end{equation}
In momentum space we obtain
\begin{equation}
\Delta_T^{ij} \left( k_0, {\mathbf k} \right) = \sum_{\lambda= \pm} \frac{1}{2\omega ^\lambda({\mathbf k}) }{\left( {\frac{{%
\varepsilon^i ( {\lambda ,{\mathbf k}})\, {\varepsilon^j}^* ( {%
\lambda ,{\mathbf k}})}} {{\left( {k_0 - \omega^\lambda({\mathbf
k}) + i\varepsilon} \right)}} - \frac{{\varepsilon^j({\lambda , -
{\mathbf k}})\, {\varepsilon^i}^* ( {\lambda , - {\mathbf k}} )}}
{{ \left( {k_0 + \omega ^\lambda({\mathbf k}) - i\varepsilon }
\right)}}} \right)}.
\end{equation}
The required expressions  for  the polarization vectors
contributions  are explicitly given by \begin{equation}
\varepsilon ^{i}(\lambda ,{\mathbf k})\, \varepsilon ^{j\ast
}(\lambda ,{\mathbf k})=\frac{1}{2}\left[ \delta
_{ij}-\frac{k_{i}k_{j}}{|{\mathbf k}|^{2}}\right] -\lambda
\frac{i}{2}\left[ \epsilon _{ijm}\frac{k_{m}}{|{\mathbf
k}|}\right].
\end{equation}
\section{Final comments}
We have constructed a perturbative modification to the classical
version of the photon-fermion  sector of the Myers-Pospelov model,
which has been subsequently quantized at the non-interacting
level. The LIV terms, which in general are of the HOTD character,
are assumed to represent very small perturbations over standard
QED. In this way, the resulting interacting quantum theory will be
required to reproduce the Lorentz covariant results of standard
QED in the limit when the LIV parameters go to zero. That is to
say, the quantum model  should be able to smoothly interpolate
between the initial Lorentz violating theory and the final Lorentz
preserving one. Such a realization already exists in the
literature for theories characterized by a fully dimensionless
Lorentz violating parameter \cite{ALFARO}. In our case Lorentz
invariance violation is characterized by the energy scale $M$, so
that these methods would not be directly applicable.

To perform the free field  quantization we have started at the
classical level using the method of Ref. \cite{CHENGETAL}, which
allowed us to successfully deal with two of the general problems
originating from the HOTD character of the terms describing the
Lorentz violations in the model: (i) the increase in the number of
degrees of freedom, which we must not allow in a perturbative
modification of standard QED, and (ii) the appearance of
Hamiltonians which are not bounded from below, which do not
provide a good starting point for quantization. The calculation of
interacting processes is deferred for a future publication.

Some  remarks regarding the effective character of the model,
together with the associated characteristic energy scales are now
in order. The combinations of parameters  $\xi/M, \eta_{1,2}/M $,
denoted collectively by $\Xi / M $,  appearing in Eqs. (2.1-2.2)
are considered as remnants of a more fundamental quantum gravity
(QG) theory, which include  effects that make space no longer
describable in terms of a continuum. Such parameters  could arise
in the process of calculating expectation values of well defined
QG operators in semiclassical states that describe Minkowski
space-time, for example, which would be necessary to derive the
induced corrections to standard particle dynamics  at low
energies. Let us emphasize that what is bounded by experiments or
observations is the ratio $\Xi / M $, so that a neat separation of
the scale $M$ and the correction coefficients $\Xi$, that  could
even  be zero if no corrections arise, is not possible  until a
semiclassical calculation is correctly performed starting from a
full quantum theory. Initially, the naive expectation was that
taking $M=M_{Planck}$ will be consistent with $\Xi $ values of
order one, which is certainly not the case. Nevertheless, we
should not rule out rather unexpected values of $\Xi$ or $M$ until
the correct calculation is done.

Let us assume that we have identified the correct separation in
${\Xi_{QG}}/{M_{QG}}$ consistent with the experimental bounds and
arising from a correct semiclassical limit of the QG theory. Then
we will interpret $ M_{QG}$  as the scale in  which quantum
effects are manifest and where space is characterized by strong
fluctuations forbidding its description as a continuum.

Nevertheless, another scale $\bar{M}$ naturally arises in this
approach, which is the one that separates the continuum
description of space from a foamy description related to quantum
effects. That is to say, for probe energies $E < \bar{M}$ we are
definitely within the standard continuum description of space
where EFT methods should apply. For probe energies $E > \bar{M} $
we enter the realm of quantum gravity and there we assume that any
EFT has to be replaced by an alternative description. It is
natural that a very large number of the basic quantum cells of
space characterized by the scale $\left( 1/M_{QG}\right)^{3}$ will
contribute to
the much larger cells characterizing a continuum description, so that we expect $\bar{M}%
\ll M_{QG}$.

The maximum allowed momenta $|{\mathbf k}_{max}|\approx
M_{QG}/\Xi_{QG}$ in the theory will be mathematically dictated by
the positivity of the normal modes energies (\ref{PHYSNEGEN1}),
(\ref{FOTFREQ}) and  certainly constitutes an extrapolation of the
EFT that can be considered as the analogous of taking the maximum
momentum equal to infinity in the standard QED case.

Nevertheless, we need to introduce and additional suppression of
the excitation modes in our EFT  which  will be settled by the
scale $\bar{M}$, thus defining the effective energy range of the
model. This is required by  the EFT description of excitations in
space which demands that the Compton wave length $1/|{\mathbf k}|$
of the allowed excitations  be larger than the scale $1/{\bar M}$
setting the onset of the continuum. The implementation of this
proposal is directly related with our demand that the quantum
model constructed from the MP theory be such that it produces a
continuous interpolation between those physical results including
$\Xi$ corrections  and those predicted by standard QED ($\Xi=0$).
Preliminary calculations of radiative corrections indicate that
such suppression, together with the desired limit,  can be
achieved by the following prescription: (i) even though the loop
integrals are all finite in the adopted MP setting, for each would
be divergent integral in the limit $\Xi \rightarrow 0$ we
introduce the
required covariant Pauli-Villars type factor with mass parameter $\bar{M}%
 \gg m^{2}$
\begin{equation}
\frac{1}{k^{2}-m^{2}}\rightarrow \frac{1}{k^{2}-m^{2}}-\frac{1}{k^{2}-\bar{M}%
^{2}}\rightarrow \frac{1}{k^{2}-m^{2}}\left( \frac{\bar{M}^{2}}{\bar{M}%
^{2}-k^{2}}\right).
\end{equation}
Besides providing a cutoff for the excitation modes in the region
${\bar M} < |{\mathbf k}|< {M_{QG}}/{\Xi _{QG}}$,  further
motivation for such factor is that it would correspond to an
adequate smooth regulator when $\Xi \rightarrow 0$. (ii) Since the
relevant scales are such that $\bar{M} \ll {M_{QG}}/{\Xi _{QG}}$,
the correct limit to standard QED will be defined by first taking
$\Xi _{QG}/M_{QG}\rightarrow 0 $, i.e. $|\mathbf{k}_{\max
}|\rightarrow \infty$ and subsequently $\bar{M}\rightarrow \infty
$. A renormalization prescription consistent with this proposal
needs to be implemented at the level of the MP model.

The work of CMR, LU and JDV has been partially supported by the
projects DGAPA-UNAM \# IN109107, CONACYT \# 47211-F and CONACYT \#
55310.

\end{document}